\documentclass[prl,twocolumn,showpacs,preprintnumbers,amsmath,amssymb]{revtex4}

\usepackage{graphicx}
\usepackage{dcolumn}
\usepackage{bm}

\begin{document}

\preprint{PRL}

\title{The spectral theorem of many-body Green's function theory when
there are zero eigenvalues of the matrix governing the
equations of motion.} 

\author{P. Fr\"obrich}
\altaffiliation[Also at ]{Institut f\"ur Theoretische Physik,
Freie Universit\"at Berlin,
Arnimallee 14, D-14195 Berlin, Germany.}

\email{froebrich@hmi.de}

\author{P.J. Kuntz}
\email{kuntz@hmi.de}
\affiliation{Hahn-Meitner-Institut Berlin,
Glienicker Stra{\ss}e 100, D-14109 Berlin,
Germany}


\date{\today}

\def\K{\mathord{\cal K}}

\def\la{\langle}

\def\ra{\rangle}

\def\ltsim{\mathop{\,<\kern-1.05em\lower1.ex\hbox{$\sim$}\,}}

\def\gtsim{\mathop{\,>\kern-1.05em\lower1.ex\hbox{$\sim$}\,}}

\begin{abstract}
In using the spectral theorem of many-body Green's function theory in order to
relate correlations to commutator Green's functions, it is necessary
in the standard procedure to consider
the anti-commutator Green's functions as well whenever the matrix governing the
equations of motion for the commutator Green's functions has zero eigenvalues.
We show that a
singular-value decomposition of this matrix allows one to reformulate the
problem in terms of a smaller set of Green's functions with an associated matrix
having no zero eigenvalues, thus eliminating the need for the anti-commutator
functions. The procedure is quite general and easy to apply. It is illustrated
for the field-induced reorientation of the magnetization of a
ferromagnetic Heisenberg monolayer and it is expected to work for more
complicated cases as well.

\end{abstract}





\maketitle

Many-body Green's function (GF) techniques make use of the spectral theorem
to connect correlation functions to the Green's functions, which
in turn are determined from equations of motion that can be
formulated conveniently in terms of a matrix equation.
Should this matrix have zero
eigenvalues, particular care must be taken, since they cannot be handled
directly by the spectral theorem for the commutator GF, since it would
require a division by zero. The standard treatment of this case demands in
addition to the commutator GF a knowledge of the anti-commutator GF (see
references \cite{ST65,RG71} and textbooks \cite{Nol86,GHE01}).
An equivalent treatment in terms of
the anti-commutator GF alone may formally appear to avoid this problem, but in
practice one is faced with the same numerical difficulties.
In this paper, we show that these difficulties
can be overcome by using a singular-value decomposition (SVD) of the
equations of motion matrix to eliminate the space connected with
the zero eigenvalues (null-space), thus reducing the
size of the problem and, at the same time, making the anti-commutator GF
superfluous. That is, one uses projection matrices furnished by the SVD to
reformulate the spectral theorem in terms of a smaller number of commutator
GF's whose associated matrix has no zero eigenvalues. We show
that this procedure is generally applicable and offers an alternative to
the standard treatment; i.e. we show that a knowledge of the commutator GF's
is sufficient.

The appearance of zero eigenvalues in the matrix governing the equations of
motion is {\em not} an unusual event. We have met this problem in much of
our previous work on the application of many-body GF's to the description
of thin ferromagnetic films \cite{EFJK99,FJK00,FKS02,HFKTJ02,FJKE00,FK02}.
We treated the
Heisenberg ferromagnet including single-ion anisotropy
and exchange anisotropy for
several values of the spin $S$ for single-layer and multi-layer systems. A
variety of decoupling approximations \cite{Tya59,AC64} to arrive
at a closed set of equations of motion were considered.
In most of these cases, the equations of motion matrix
has at least one zero eigenvalue per layer and, if higher-order Green's
functions \cite{FKS02} are used to achieve an exact treatment of the single-ion
anisotropy, there are two zero eigenvalues per layer. Calculation of
many-layer systems therefore leads to matrices with a large number of zero
eigenvalues. This, coupled with the fact that the matrices are in general
nonsymmetric, leads in some cases to numerical difficulties which can
best be circumvented by eliminating
the null-space. The method for doing this
may be considered as an alternative to the
standard procedure of handling zero eigenvalues. In the following, we first
introduce our methodology and notation by reviewing briefly the standard
text-book treatment of zero eigenvalues,
then we describe the new method which is based on the singular value
decomposition of the matrix governing the equation of motion, and finally,
by way of example, we apply the new method to
the field-induced reorientation of
the magnetization of a ferromagnetic Heisenberg monolayer.


{\em The standard procedure.}$-$The equation of motion for a
Green's function vector
\begin{equation}
{\bf G}_{\eta,\omega}=\la\la A;B\ra\ra_{\eta,\omega}
\label{1}
\end{equation}
in energy space is (omitting the subscript $\omega$)
\begin{equation}
\omega\la\la A;B\ra\ra_\eta=\la [A,B]_\eta\ra+\la\la [A,H]_-;B\ra\ra_\eta,
\label{2}
\end{equation}
where  $\la ... \ra=Tr(...e^{-\beta H})/Z$ is the thermodynamic expectation
value, $H$ is the Hamiltonian and  $A$ and $B$ are operators depending on
the problem under consideration, and $\eta=\mp 1$ refers to the commutator or
anti-commutator Green's functions, respectively.

Repeated application of equation~(\ref{2})
to the higher-order Green's functions
appearing on the right hand side, $\la\la [A,H]_-;B\ra\ra$,
results in an infinite hierarchy of equations.
In order to obtain a set of solvable equations, one has to terminate
this hierarchy, usually by expressing the Green's functions of a
particular order in terms of those of lower order. In many
applications, only the lowest-order functions are retained using
a decoupling procedure of the form
\begin{equation}
\la\la [A,H]_-;B\ra\ra_\eta\simeq {\bf \Gamma} \la\la A;B\ra\ra_\eta\ ,
\label{3}
\end{equation}
where $\bf \Gamma$ is the (in general {\em non-symmetric})
matrix expressing the higher-order Green's
functions in terms of linear combinations of lower-order ones.
In compact matrix notation, the equation of motion is
\begin{equation}
(\omega {\bf 1}-{\bf \Gamma}) {\bf G}_\eta = {\bf A}_\eta,
\label{4}
\end{equation}
where ${\bf 1}$ is the unit matrix, and the inhomogeneity vector is
${\bf A}_\eta =\la [A,B]_\eta\ra$.

It is now convenient to
introduce the notation of the eigenvector method of
reference \cite{FJKE00}, since it is
particularly suitable for the multi-dimensional problems in which
many zero eigenvalues are likely to appear.
One starts by diagonalizing the  matrix
${\bf \Gamma}$
\begin{equation}
{\bf L\Gamma R}={\bf \Omega},
\label{5}
\end{equation}
where $\bf \Omega$ is the diagonal matrix of $N$ eigenvalues,
$\omega_\tau\  ({\tau=1,..., N})$, $N_0$ of which are zero and
$(N-N_0)$ are non-zero.
 The matrix ${\bf R}$ contains the right eigenvectors as columns and its
inverse
${\bf L}={\bf R}^{-1}$ contains
the left eigenvectors as rows. ${\bf L}$ is
constructed such that ${\bf LR}={\bf RL}={\bf 1}$.
Multiplying equation (\ref{4}) from the left by
${\bf L}$, inserting ${\bf 1}={\bf RL}$, and defining new vectors
${\cal G}_\eta={\bf LG}_\eta$ and ${\cal A}_\eta={\bf LA}_\eta$
one finds
\begin{equation}
(\omega{\bf 1}-{\bf \Omega}){\cal G}_\eta={\cal A}_\eta.
\label{6}
\end{equation}
Each of the components $\tau$ of this Green's function vector
has but a single pole
\begin{equation}
{(\cal G}_\eta)_\tau=\frac{({\cal A}_\eta)_\tau}{\omega-\omega_\tau}\ .
\label{7}
\end{equation}
This allows the spectral theorem, (see e.g.
\cite{Nol86,GHE01}), to be applied
to each component of the Greens's function vector {\em separately}:
Define the correlation vector
${\cal C}={\bf LC}$, where ${\bf C}=\la BA\ra$ is the vector of correlations
associated with ${\bf G}_\eta$. From the spectral theorem, the
component $\tau$ for the anti-commutator case ($\eta=+1$) is
\begin{equation}
{\cal C}_\tau=\frac{({\cal A}_{\eta})_\tau}{e^{\beta\omega_\tau}+\eta}\ .
\label{8a}
\end{equation}
Use of the anti-commutator functions appears at first sight more straightforward
(zero eigenvalues seem to make no problem), but leads in general to momentum
dependent inhomogeneities, which are difficult to treat.
However, it is usually easier to employ
commutator Green's functions, because in this case the inhomogeneity vectors
are momentum independent.

For the commutator functions ($\eta=-1$),
the $N-N_0$ components for $\omega_\tau \neq 0$, ${\cal C}^1$ (the upper index
1 refers to the non-null space) , are
\begin{equation}
({\cal C}^1)_\tau=\frac{({\cal A}_\eta)_\tau}{e^{\beta\omega_\tau}+\eta}\ .
\label{8c1}
\end{equation}
This equation cannot be used to define
the $N_0$ components ${\cal C}^0$
corresponding to $\omega_\tau=0$ because of the zero in the denominator.
Instead, one must enlist the help of the {\em anti-commutator} Green's function:
\begin{equation}
({\cal C}^0)_{\tau_0}=\lim_{\omega\rightarrow 0}\frac{1}{2}\omega
({\cal G}_{\eta=+1})_\tau.
\label{8c0}
\end{equation}
The components of ${\cal C}^0$, indexed by $\tau_0$ (the upper index 0 refers
to the null space), can be simplified by using the relation between the
commutator and anti-commutator inhomogeneities,
${\bf A}_{1}={\bf A}_{-1}+2{\bf C}$. This yields, for
$\omega_{\tau o}=0$,
\begin{eqnarray}
({\cal C}^0)_{\tau_0}&=&\frac{1}{2}\lim_{\omega\rightarrow 0}
\frac{\omega({\cal A}_{+1})_{\tau_0}}{\omega-\omega_{\tau_0}} =
\frac{1}{2}({\cal A}_{+1})_{\tau_0}\nonumber\\
&=&\frac{1}{2}({\bf L}^0({\bf A}_{-1}+2{\bf C}))_{\tau_0}
=({\bf L}^0{\bf C})_{\tau_0},
\label{9}
\end{eqnarray}
where use has been made of the relation ${\bf L}^0{\bf A}_{-1}=0$, which
derives from the fact that the
commutator Green's function is regular at the origin \cite{FJKE00}.

At this point it is again convenient to introduce superscripts 0 and 1 to
denote vectors corresponding to zero and non-zero eigenvalues,
respectively. The right and left eigenvectors and the correlation vectors may
then be partitioned as ${\bf R} = ({\bf R}^1 \ {\bf R}^0)$ and
\begin{equation}
   {\bf L} = \left( \begin{array}{c}
                         {\bf L}^1  \\ {\bf L}^0
                                           \end{array}      \right)\ \ \ \ \ \
                   {\cal C} = \left( \begin{array}{c}
                         {\cal C}^1  \\ {\cal C}^0
                    \end{array}      \right)\ ,
\end{equation}
where the correlation vectors from equations~(\ref{8c1}) and~(\ref{9}) are then
${\cal C}^0 = {\bf L}^0{\bf C}$
and ${\cal C}^1 = {\cal E}^1 {\bf L}^1 {\bf A}$,
and ${\cal E}^1$ is the $(N-N_0)\times(N-N_0)$ matrix with
$1/(e^{\beta\omega_\tau}-1)\ $
on the diagonal.

Multiplying the correlation vector $\cal C$ from the left by $\bf R$ yields a
compact matrix equation for the original correlation vector $\bf C$:
\begin{equation}
{\bf C}={\bf R}^1{\cal E}^1{\bf L}^1{\bf A}+{\bf R}^0{\bf L}^0{\bf C},
\label{11}
\end{equation}

This equation is in momentum space
and the coupled system of integral equations obtained by  Fourier
transformation to real space has to be solved self-consistently. This
was done in all our previous work \cite{EFJK99,FJK00,FKS02,HFKTJ02,FJKE00,FK02}.

{\em The singular value decomposition.$-$}
We now present an alternative
method of treating the zero eigenvalues. Starting again from
the equation of motion~(\ref{4}), we write the ${\bf \Gamma}$ matrix
as a singular value decomposition \cite{PFTV89}:
\begin{equation}
{\bf \Gamma}={\bf UW}\tilde{{\bf V}}.
\label{12}
\end{equation}
The matrix  ${\bf W}$ is a diagonal matrix whose elements are the
singular values, which are $\geq 0$ and ${\bf U}$ and ${\bf V}$ are
orthogonal matrices: $\tilde{\bf U}{\bf U}={\bf 1}$ and
$\tilde{\bf V}{\bf V}={\bf 1}$, where $\tilde{\bf V}$ denotes the
transpose of ${\bf V}$. ${\bf U}$, ${\bf V}$ and $\bf W$ can be obtained
very efficiently numerically \cite{PFTV89}. However, one can also
get the vectors ${\bf V}$ and ${\bf U}$ directly
by diagonalizing $\tilde{\bf \Gamma}{\bf \Gamma}$ or
${\bf \Gamma}\tilde{\bf \Gamma}$ respectively.  The singular values are just
the positive square roots of the eigenvalues of these matrices, ${\bf W}^2$.
\begin{eqnarray}
\tilde{\bf V}\tilde{\bf \Gamma}{\bf \Gamma}{\bf V}&=&\tilde{\bf V}{\bf V}{\bf
W}\tilde{\bf U}{\bf U}{\bf W}\tilde{\bf V}{\bf V}={\bf W}^2\ , \nonumber\\
\tilde{\bf U}{\bf \Gamma}\tilde{\bf \Gamma}{\bf U}&=&\tilde{\bf U}{\bf U}{\bf
W}\tilde{\bf V}{\bf V}{\bf W}\tilde{\bf U}{\bf U}={\bf W}^2\ .
\label{13}
\end{eqnarray}
Note that the ${\bf \Gamma}$ matrix is fully determined by the
non-zero singular values and the corresponding eigenvectors:
\begin{equation}
{\bf \Gamma}={\bf UW}\tilde{\bf V}={\bf uw}\tilde{\bf v}.
\label{14}
\end{equation}
$\bf u$ and ${\bf v}$ are $N\times (N-N_0)$ matrices
obtained from $\bf U$ and $\bf V$ by omitting the columns corresponding
to the zero singular values.
The matrix ${\bf w}$ is the $(N-N_0)\times(N-N_0)$ diagonal matrix
with the non-zero singular values on the diagonal.

The crucial point is that the dimension of the equations of motion can
be reduced by the number of zero singular values by applying the following
transformations
\begin{eqnarray}
{\bf \gamma}&=&\tilde{\bf v}{\bf \Gamma}{\bf v},\nonumber\\
{\bf g}&=&\tilde{\bf v}{\bf G}, \nonumber\\
{\bf a}&=&\tilde{\bf v}{\bf A},
\label{15}
\end{eqnarray}
and realizing that ${\bf \Gamma}={\bf \Gamma}v\tilde{v}$ and $\tilde{v}v=1$.

The $(N-N_0)$-dimensional equation of motion then becomes
\begin{equation}
(\omega{\bf 1}-{\bf \gamma}){\bf g}={\bf a}.
\label{16}
\end{equation}
Here again the eigenvector method can be used, where ${\bf l}$ and ${\bf r}$
diagonalize the ${\bf\gamma}$-matrix, ${\bf l\gamma r}=\omega^1$. Applying the
spectral theorem to the equation of motion with the
matrix ${\bf \gamma}$ (which now has no zero eigenvalues) yields the equation
for the correlations ${\bf c}=\tilde{\bf v}{\bf C}$ in momentum space
\begin{equation}
{\bf c}={\bf r}{\cal E}^1{\bf la }
\label{17}
\end{equation}
where the $(N-N_0)\times(N-N_0)$ diagonal matrix ${\cal E}^1$ has the same
elements as before,
$1/(e^{\beta\omega_\tau}-1)$.
In order to determine the correlations in coordinate space,
one has to perform a Fourier transform and then solve self-consistently the
system of integral equations
\begin{equation}
0=\int d{\bf k}({\bf r}{\cal E}^1{\bf l}\tilde{\bf v}{\bf A}-\tilde{\bf v}{\bf
C})\ ,
\label{18}
\end{equation}
where ${\bf A}$ and ${\bf C}$ are the inhomogeneity and correlation vectors of
the original problem, and the integration is over the first Brillouin zone.

The advantages of the new procedure are that, after
projecting onto
the non-null space, it is no longer necessary to calculate the term in the
spectral theorem which is connected with the anti-commutator Green's function,
and the dimension of the final system of equations is reduced by the number of
zero eigenvalues of the original $\bf \Gamma$ matrix.

{\em Example: The Heisenberg monolayer.}$-$ As an example, we apply the
new procedure to a
ferromagnetic Heisenberg monolayer with spin $S=1$, in which the magnetization
can be reoriented by applying an external field. This is an example in which
zero eigenvalues of the ${\bf \Gamma}$ matrix occur and is
treated in references \cite{FJK00,FJKE00} with the standard procedure not only
for the $S=1$ monolayer but also for the multilayer case and for $S\geq 1$.

We restrict ourselves to a spin Hamiltonian  consisting of an
isotropic Heisenberg exchange
interaction with strength $J_{kl}$ between nearest neighbour lattice sites,
 a second-order single-ion lattice
anisotropy with strength $K_{2,k}$,  and an external magnetic
field ${\bf B}=(B^x,B^y,B^z)$:
\begin{eqnarray}
{\cal
H}=&-&\frac{1}{2}\sum_{\la kl\ra}J_{kl}(S_k^-S_l^++S_k^zS_l^z)
-\sum_kK_{2,k}(S_k^z)^2\nonumber\\
&-&\sum_k\Big(\frac{1}{2}B^-S_k^++\frac{1}{2}B^+S_k^-+B^zS_k^z\Big).
\label{22}
\end{eqnarray}
Here the notation $S_k^{\pm}=S_k^x\pm iS_k^y$ and $B^{\pm}=B^x\pm iB^y$ is
introduced, where $k$ and $l$ are lattice site indices and $\la kl\ra$
indicates summation over nearest neighbours only.

%
%

Applying a Tyablikov (RPA) decoupling to the exchange interaction term, and an
Anderson-Callen decoupling to the single-ion anisotropy term one obtains
\cite{FJK00,FJKE00}
 equations of motion for the monolayer of the form
\begin{equation}
(\omega{\bf 1}-{\bf \Gamma}){\bf G}_\eta={\bf A}_\eta
\label{24}
\end{equation}
with the Green's function
and inhomogeneity vectors
\begin{equation}
{\bf G}_\eta^{mn}({\bf{k},\omega})\  =
\left( \begin{array}{c}
G_\eta^{+,mn}({\bf{k}},\omega) \\ G_\eta^{-,mn}({\bf{k}},\omega)  \\
G_\eta^{z,mn}({\bf{k}},\omega)
\end{array} \right), \hspace{0.5cm}
{\bf A}_\eta^{mn} {=}
 \left( \begin{array}{c} A_\eta^{+,mn} \\ A_\eta^{-,mn} \\
A_\eta^{z,mn} \end{array} \right) \;,
\label{25}
\end{equation}
where
$G_{ij,\eta}^{\alpha,mn}(\omega)=\la\la
S_i^\alpha;(S_j^z)^m(S_j^-)^n\ra\ra_{\omega,\eta}$
and $\alpha=(+,-,z)$ takes care of all directions in space.
For a field in the $x$-direction only, ${\bf B}=(B^x,0,0)$, the
reorientation of the magnetization takes place in the $xz$-plane, and the
${\Gamma}$ matrix for spin $S=1$ ($n\geq 1, m\geq 0, m+n\leq 2)$ is
\begin{equation}
 {\bf \Gamma}= \left( \begin{array}
{@{\hspace*{3mm}}c@{\hspace*{5mm}}c@{\hspace*{5mm}}c@{\hspace*{3mm}}}
\;\;\;H^z & 0 & -H^x \\ 0 & -H^z & \;\;\;H^x \\
-\frac{1}{2}{H}^x & \;\frac{1}{2}{H}^x & 0
\end{array} \right)\ .
\label{26}
\end{equation}
where
\begin{eqnarray}
H^z&=&Z+\la S^z\ra J(q-\gamma_{\bf k})\ ,
\nonumber\\
Z&=&K_{2}2\la S^z\ra
\Big(1-\frac{1}{2S^2}[S(S+1)-\la S^zS^z\ra]\Big)\ ,
\nonumber \\
H^x&=&B^x+\la S^x\ra J(q-\gamma_{\bf k})\ .
\label{27}
\end{eqnarray}
For a square lattice with lattice constant unity,  $\gamma_{\bf
k}=2(\cos k_x+\cos k_y)$, and $q=4$ is the
number of intra-layer nearest neighbours.
\begin{figure}
\begin{center}
\protect
\includegraphics*[bb=80 90 540 700,
angle=-90,clip=true,width=8cm]{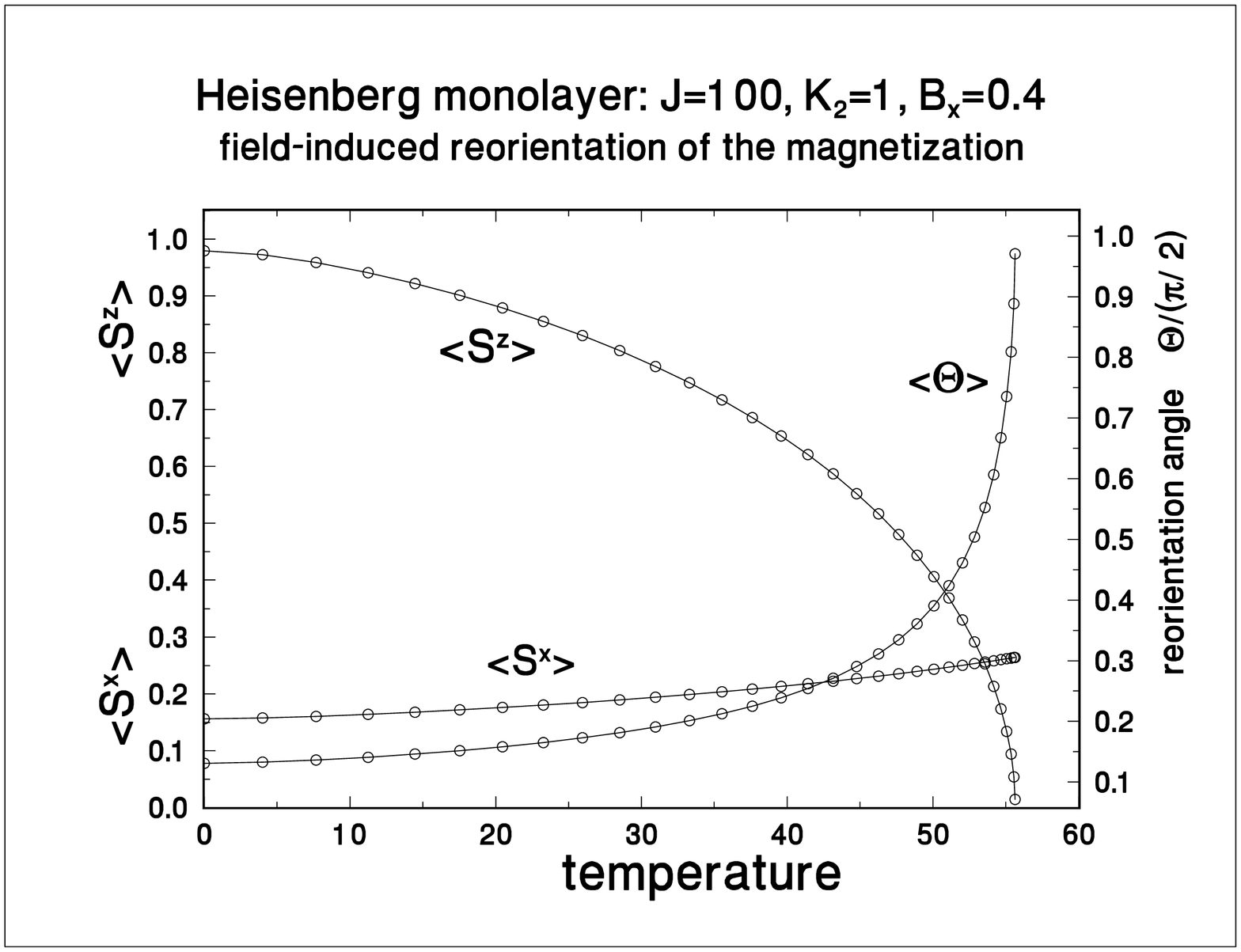}
\protect
\caption{Magnetizations $\la S^z\ra$ and $\la S^x\ra$ and the reorientation
angle $\Theta$ for a spin S=1 Heisenberg monolayer as function of the
temperature.}
\end{center}
\end{figure}
%
The singular value decomposition of equation (\ref{12}) can be performed
analytically for the monolayer case.
%
%
The singular values of the diagonal matrix $W$ are given as the positive
square roots of the eigenvalues of the matrix
${\bf \Gamma}\tilde{\bf \Gamma}$ or $\tilde{\bf \Gamma}{\bf \Gamma}$
\begin{equation}
 {\bf W}=\left( \begin{array}
 {@{\hspace*{3mm}}c@{\hspace*{5mm}}c@{\hspace*{5mm}}c@{\hspace*{5mm}}}
 \;\;\;\epsilon_1 & 0 &\ 0 \\ 0 &
 \epsilon_2 & \ 0\\ 0& 0 &\ 0
 \end{array} \right)\ ,
\label{29}
\end{equation}
where $\epsilon_1=\sqrt{H^zH^z+2H^xH^x}$,
$\epsilon_2=\sqrt{H^zH^z+\frac{1}{2}H^xH^x}$, and one singular value zero.
The matrices ${\bf U}$ and ${\bf V}$ are obtained from the eigenvectors of
these matrices%
\begin{equation}
{\bf U}=\left( \begin{array}
{@{\hspace*{3mm}}c@{\hspace*{5mm}}c@{\hspace*{5mm}}c@{\hspace*{5mm}}}
\;\;\;\frac{-\sqrt{2}}{2} &\frac{-H^z}{\sqrt{2}\epsilon_2} &\
\frac{H^x}{2\epsilon_2} \\ \frac{\sqrt{2}}{2} &
\frac{-H^z}{\sqrt{2}\epsilon_2}
& \ \frac{H^x}{2\epsilon_2}\\ 0& \frac{H^x}{\sqrt{2}\epsilon_2} &\
\frac{H^z}{\epsilon_2} \end{array} \right)\ .
\label{30}
\end{equation}
and
\begin{equation}
\tilde{\bf V}=\left( \begin{array}
{@{\hspace*{3mm}}c@{\hspace*{5mm}}c@{\hspace*{5mm}}c@{\hspace*{5mm}}}
\;\;\; \frac{-H^z}{\sqrt{2}\epsilon_1} &  \frac{-H^z}{\sqrt{2}\epsilon_1} &\
\frac{\sqrt{2}H^x}{\epsilon_1} \\ \frac{-\sqrt{2}}{2} &
\frac{\sqrt{2}}{2} & \ 0\\ \frac{H^x}{\epsilon_1}& \frac{H^x}{\epsilon_1} &\
\frac{H^z}{\epsilon_1} \end{array} \right)\ .
\label{31}
\end{equation}
The lower case rectangular matrices ${\bf u}$ and $\tilde{\bf v}$
introduced above are obtained by omitting the last column in ${\bf U}$
and the last row in $\tilde{\bf V}$ respectively.
One sees explicitly that $\tilde{\bf v}{\bf v}={\bf 1}$ and that
${\bf v}\tilde{\bf v}$ is a projection operator onto the non-null space
because it has two eigenvalues equal to $1$ and one equal to $0$, and
$({\bf v}\tilde{\bf v})^2={\bf v}\tilde{\bf v}$.

Having constructed the matrix $\tilde{\bf v}$, we now solve equation (\ref{18})
for the spin $S=1$ monolayer
with the curve following method, described in detail in reference
\cite{FKS02}. The result, shown in figure 1,
agrees perfectly with that obtained
in references \cite{FJK00,FJKE00} obtained with the standard procedure,
i.e. by solving equation~(\ref{11}). In that work,
spins $S>1$ and multilayer systems were also considered. This can also be
handled by the new algorithm but, in this case, the singular value
decomposition and the matrix $\tilde{\bf v}$  have to be obtained
from an appropriate numerical method \cite{PFTV89}.

{\em Summary.}$-$ We have developed a new method for dealing with
zero eigenvalues of the matrix governing the equations of motion for many-body
Green's functions by applying a singular value decomposition to this matrix.
This enables one to eliminate the null space, thus
reducing the number of integral equations which have to be solved
self-consistently by the number of zero eigenvalues of the equation of motion
matrix. The calculation of the anti-commutator Green's
function, necessary in the standard procedure using
the spectral theorem, is superfluous in the new procedure, which is
demonstrated for
the example of the field induced reorientation of the magnetization of a spin
$S=1$ Heisenberg ferromagnetic monolayer. The method is expected to work
for more complicated cases, where it has decided
advantages over the usual procedure.

\end{document}